\documentclass[sigconf]{acmart}

\usepackage{pdfpages}

\AtBeginDocument{%
  \providecommand\BibTeX{{%
    \normalfont B\kern-0.5em{\scshape i\kern-0.25em b}\kern-0.8em\TeX}}}

\copyrightyear{2021} 
\acmYear{2021} 
\setcopyright{acmcopyright}\acmConference[AIES '21]{Proceedings of the 2021 AAAI/ACM Conference on AI, Ethics, and Society}{May 19--21, 2021}{Virtual Event, USA}
\acmBooktitle{Proceedings of the 2021 AAAI/ACM Conference on AI, Ethics, and Society (AIES '21), May 19--21, 2021, Virtual Event, USA}
\acmPrice{15.00}
\acmDOI{10.1145/3461702.3462565}
\acmISBN{978-1-4503-8473-5/21/05}

\begin{document}
\fancyhead{}

\title{Explainable AI and Adoption of Financial Algorithmic Advisors: an Experimental Study}

\author{Daniel Ben David}
\email{daniel.ben-david@mail.huji.ac.il}
\affiliation{%
  \institution{The Hebrew University of Jerusalem Finance Department \\ Intuit Inc.}
  \country{}
}

\author{Yehezkel S. Resheff}
\authornote{The bulk of the research was conducted while at Intuit Inc.}
\affiliation{%
  \institution{Department of Computer Science, Holon Institude of Technology }
  \country{}
}

\author{Talia Tron}
\affiliation{%
  \institution{Intuit Inc.}
  \country{}
}

\renewcommand{\shortauthors}{Daniel Ben David, et al.}

\begin{abstract}
 We study whether receiving advice from either a human or algorithmic advisor, accompanied by five types of Local and Global explanation labelings, has an effect on the readiness to adopt, willingness to pay, and trust in a financial AI consultant. We compare the differences over time and in various key situations using a unique experimental framework where participants play a web-based game with real monetary consequences. We observed that accuracy-based explanations of the model in initial phases leads to higher adoption rates. When the performance of the model is immaculate, there is less importance associated with the kind of explanation for adoption. Using more elaborate feature-based or accuracy-based explanations helps substantially in reducing the adoption drop upon model failure. Furthermore, using an autopilot increases adoption significantly. Participants assigned to the AI-labeled advice with explanations were willing to pay more for the advice than the AI-labeled advice with a No-explanation alternative. These results add to the literature on the importance of XAI for algorithmic adoption and trust.
\end{abstract}



\maketitle

\section{Introduction}

The use of machine learning and other automation methods is becoming overwhelmingly popular and increasingly available in different facets of everyday life. From basic research these methods have made their way into medicine, transportation, business processes, retail, customer service, and diverse financial services. At times these methods are utilized to do work \textit{for} individuals, driving down costs of previously labour intensive products and services. Other times the same methods are utilized to automatically make decisions \textit{about} individuals.

When algorithms take part in decision making with significant impact to individuals, there is a burden of explainability that naturally falls on the providers of the system. In some cases there are regulations that imposes obligatory explanation of decisions as an integral part of their output (for example, the recent GDPR legislation by the European Union which requires the maker of AI algorithms which ``significantly affect'' decisions to explain how any output was obtained  \cite{ goodman2017european}). In other cases, a deep understanding of how an output was generated is crucial for human based decision making in interaction with the automatic process (e.g. in some medical, and security applications). However, in many other cases, and in various fields, the importance of explanations is first and foremost in the effect on the perceived trustworthiness of the system \cite{yin2019understanding, hoffman2018metrics,mcknight2002developing}, and hence on the readiness of consumers to adopt (RTA) the AI service, and pay for it. 

In recent years, the field of explainable AI (XAI) has seen a boost in interest from the community, with many new approaches and ideas. Most of the effort is on the technological and algorithmic side -- borrowing ideas from game theory, statistics, and machine learning, to develop fast and accurate techniques which explain black-box or opaque models \cite{molnar2019Interpretable,lundberg2017unified,kim2017interpretability}, or train AI which is more interpretable by nature \cite{angelino2017learning, lou2013accurate}. Another part of the literature deals with the independently important question of evaluating explanations based on their attributes, and with user evaluation of explanations and the consequent effect on behavior.

\subsection{Attributes of AI Explanations} 

The first important attribute when considering what explanation to generate is the type of information the explanation should convey. We find in the literature three main approaches to explanations: (1) Global Explanations, (2) Local explanations and (3) what might be called Social Influence Explanations. 

\textit{Global explanation} techniques provide an overview of what an algorithm is doing as a whole. The aim of this type of explanation is to convey to a human what the algorithm is doing rather than explain the process that lead to a specific prediction or decision. These methods often include summarized information about how a model uses features to produce prediction (some popular approaches include various notions of feature importance, dependence plots, and global Shapley values), prototype example predictions, or a simplified, interpretable approximation of a black-box model (aka surrogate models) \cite{molnar2019Interpretable,kim2017interpretability,guidotti2018survey,vilone2020explainable}. 

Another type of global explanation that is completely independent of the algorithm used is the presentation of any type of meta-information that sheds light about the process. This includes transparency around how the model was trained, the type of data that was used, or even simply reporting model performance statistics. Global explanations are for the most part less costly to produce in real-world systems, compared to the alternatives, and are readily available in most cases, making them appealing in practise. 

\textit{Local explanation} techniques, on the other hand, provide a more detailed description of how the model came up with a \textit{specific} prediction. These may include information about how a model uses features to generate a specific output \cite{lundberg2017unified, ribeiro2016should, ribeiro2018anchors,bau2017network,guidotti2018survey,vilone2020explainable}, how a perturbation in input will influence the output \cite{wachter2017counterfactual, dhurandhar2018explanations}, or a comparison of the specific input-output pair at hand to the model's output on similar input data \cite{kim2016examples}. In general, local explanation techniques are more costly in time and resources since they must be computed on a case-by-case basis rather than globally for the entire system. Furthermore, local explanations are inherently only available once the system is being used, and not applicable when the aim is to convince a user beforehand and build \textit{a-priori} trust and consent that is independent of actual experience with the system. 

\textit{Social Influence Explanations} techniques contain a type of information which relates to the way socially relevant others behave. These methods are typically discussed in the context of recommendation systems; a system using this sort of explanation may show a report on model adoption statistics, or the ranking of a specific item or of the entire system by users with a similar profile or shared characteristics \cite{tintarev2007survey,herlocker2000explaining}. We will not use this technique in our study, and it is only briefly discussed here for the completeness of the attributes review. 

An additional factor which should be taken into account is the way explanations are presented. The amount of information provided, phrasing, and choice of specific words may all affect what individuals perceive \cite{doran2017does,herlocker2000explaining}. In parallel with the textual message, the literature on explanations explores the visual interface, and which visual method is the more effective way to present a recommendation (e.g. star ranking, histograms, neighbors ranking tables, pies, word cloud) \cite{tintarev2007survey}.

With the multitude of methods that currently exist, it is not clear how to choose for a specific case. What constitutes a good explanation, and is it constant or individual and context dependant?

\subsection{Evaluating explanations}

Unlike the well defined metrics used to evaluate machine learning, such as accuracy and precision for model performance, the definition of a good explanation of the output or nature of a model is somewhat vague. The definition is presumably not universal, but rather highly dependent on what we are trying to achieve by augmenting an algorithm with an explanation. Objectives may vary dramatically for different use-cases, ranging for instance from helping the user understand how a decision was made, to attempting to convince a user to adopt the system or take a recommendation. As a result, the evaluation approaches and methodologies vary between different domains. 


Explanation evaluation can be conceptually divided into qualitative and quantitative evaluation techniques. Within the quantitative measures, one can find lines of work that focus on evaluating the mathematical (or statistical) attributes of the explanation that is generated. For example, when the explanation is presented as a report of feature importance (i.e. a list or raking of the features that were instrumental in making the decision), statistical properties such as local accuracy \cite{lundberg2017unified} help determine the quality of the list. This type of evaluation completely avoids the question of what is useful or leads to adoption and trust. In some cases the truthfulness of explanations takes precedence over all else, justifying this view. Other times it makes sense to sacrifice on this front in order to achieve the actual goal of the system.

Much of the relevant literature from the field of recommendation systems focuses on measuring usage indicators to evaluate effectiveness -- the extent to which the system assists the user in making better decisions compared to previous behaviour, and efficiency -- the extent to which it helps users make faster decisions \cite {tintarev2007survey, herlocker2000explaining}.

The qualitative measurement literature is focused mostly on user understanding and reaction to the explanations. This line of research uses various questionnaires. Many measures have been suggested to reflect the different aspects of what is important when designing explanations. These include \textit{transparency} (the level of detail provided), \textit{scrutability} (the extent to which users can give feedback to alter the AI system when it's wrong), \textit{trust} (the level of confidence in the system), \textit{persuasiveness} (the degree to which the system itself is convincing in making users buy or try recommendations given by it), \textit{satisfaction} (the level to which the system is enjoyable to use) and \textit{user understanding} (the extent a user understands the nature of the AI service offered, or alternatively the level of similarity of an explanation generated by the automatic method to explanations produced by a human being) \cite{ribeiro2016should,lombrozo2006structure,kulesza2015principles,tintarev2007survey,herlocker2000explaining}.



In our study, we attempt to form a synergy between explainable AI and the multiple fields that have previously studied algorithmic adoption, and machine-human relations. We use the quantitative measure of Readiness to Adopt (RTA), simply defined as the fraction of users that use the AI system when presented with the choice, and later the Willingness to Pay (WTP) to explore acceptance and its relations to trust and user satisfaction. We study the impact of the explanation type (both global and local) on the adoption of and payment willingness towards an AI financial decision-making advisor. We do so in a unique experimental framework, in a controlled environment with real money consequences, and repeating interactions which evolve over time. In addition, we relate the Readiness To Adopt (RTA) and the Willingness to Pay (WTP) in the different treatments to known constructs from the literature on trust \cite{gefen2003trust,mcknight2002developing,komiak2006effects}, and perceived quality of explanations \cite{komiak2006effects,hoffman2018metrics}
. This innovative framework allows us to examine whether there are differences in adoption and payment when the advice is labeled as human advisor compare to AI based algorithm, what are the effects of types of explanations on initial adoption, adoption over time, algorithm aversion when the model fails, fully autonomous decision-making services, and willingness to pay. The research questions we aim to explore include: whether different types of explanations are important for adoption of an financial AI algorithm, when all else is equal (actual advice is the same)? Will providing more detailed information (via local explanations) increase trust, RTA? How will this effect change after multiple interactions with the model? How will a failure in model performance effect the way people perceive financial algorithms with different explanations? Do explanations influence the RTA autonomous advice or have an effect when the advice is costly? Do explanations influence the consumer of the advice WTP?

To the best of our knowledge, this study is the first to compare the effect of different types of textual explanations of AI advice in term of RTA over time and in different situations. Our approach stems from the conjecture that the model-consumer interaction has a dynamic nature, which is reference dependent, and it may be influenced by various factors such as familiarity, past performance, and potential cost. This implies that one explanation can be optimal to make a good sense of trustworthiness, and gain initial trust and understanding, while a different one could be better fitting after a period of using the model, or in a such a case when it fails.
In addition, and to our knowledge, this study is the first that explores the affect of explanation on customers Willingness to Pay (WPT) for the AI advice.

The rest of the paper is organized as follows: in the next section we provide a detailed description of experiment methodology and design, we discuss the choice of experimental treatments and game flow and relate it to the existing literature from the field of explainable AI and algorithmic trust respectively. Next, we present the results from an online experiment showing how explanations affect adoption and trust in the different phases of human-AI interactions, as well as the consequences for participants' willingness to pay for the service. Finally, we discuss the finding from this study, and suggest future directions with the potential to broaden the scope of the current work and generalize to other domains.

\section{Methods}

\begin{table*}[h]
\centering

\begin{tabular}{{p{4cm}p{8cm}}}
\hline
Condition & Explanation Type \\
\hline
\textit{Human Expert} & The human advisor recommendation is to make 6 cups of lemonade. \\
\textit{No explanation} & The algorithm recommendation is to make 6 cups of lemonade. \\
\textit{Global explanation} & Based on data from lemonade stands over several years, the algorithm recommendation is to make 6 cups of lemonade. \\
\textit{Feature-based} & Based on data from lemonade stands over several years, your previous sales, and market demand, the algorithm recommendation is to make 6 cups of lemonade. \\
\textit{Performance-Based} & Based on data from lemonade stands over several years, the algorithm recommendation, with 90 percent certainty, is to make 6 cups \\

\hline
\end{tabular}

  \caption{The text presented to users in each experimental treatment. The recommended number presented changed by the specific day advice and varied between 1-10}
  \label{tab:exp-conditions}
\end{table*}

\subsection{Study Design}
The study consistent of a 3 parts as follow:
\begin{enumerate}
    \item Pre-game quantitative questionnaire.
    participants had a time limit of 3 minutes to answer 3 simple mathematical questions (addition and subtraction). Upon completing this part, participants earned 20 initial game coins that were used in the main part of the study. Each game coin during the game was worth 2 U.S. cents. The purpose of this stage was twofold: The first was to ensure participants' attention during the experiment. The second relates to the mental accounting literature \cite{henderson1992mental}. We wanted participants to treat the game coins account as money that they earned while investing effort, rather than money that they obtained as a "reward".
    \item The main part of the study. A fun, interactive, decision-making game -- The Lemonade Stand (see full description below). In this part participants could gain more coins depending on their decisions, and potentially use an advisor, that was labeled differently with several explanation, in doing so.
    \item A post-game questionnaires about trust, engagement, explanation satisfaction and personal demographic details \footnote{The questionnaires about Trust and explanation satisfaction are based on  \cite{gefen2003trust,komiak2006effects,mcknight2002developing,hoffman2018metrics}. The metrics and the questionnaires can be found in appendix A.}. 
    
\end{enumerate}

Figure \ref{fig:game_flow} illustrates the lemonade stand game. In the beginning of the game participants were instructed as followed: ``You own a lemonade stand, your goal is to make as much money as you can in 2 weeks by selling lemonade. Decide how many cups you want to make, per day, based on the price of lemons and the weather forecast.'' At the beginning of each day, a weather forecast (sunny, cloudy, or rainy) was displayed together with the varying price of lemons (0.45-0.55 coins per cup). The lemonade selling price was fixed (1 coin). Participants had to decide how many cups of lemonade they wanted to manufacture (1 - 10 cups) based on the information provided to them. At the end of each day, participants were given a short summary of the actual weather that occurred, the demand for lemonade (in cups), and how much they were able to sell (the minimum of the number produced and the demand). In case more lemonade was made than the actual demand that day, cups were not rolled over to the next day. 

The weather for each day was sampled uniformly for each participant, and the weather forecast was generated randomly such that the forecast was accurate 80\% of the time (and uniformly one of the remaining two options the rest of the time). The demand for lemonade was only a (albeit stochastic) function of the actual weather each day (see table \ref{tab:weather-demand}) for ranges.

\begin{table}[!h]
   \centering

\begin{tabular}{p{3cm}p{3cm}}
\hline
Weather & Demand \\
\hline
Rainy & 0-2 \\
Cloudy & 3-7 \\
Sunny & 8-10 \\ 
\hline
\end{tabular}

    \caption{The ranges of demand level conditioned on weather. The demand for lemonade cups is sampled uniformly from the range associated with the actual weather each day.}
    \label{tab:weather-demand}
\end{table}

\subsection{Game flow}

\begin{figure*}[t]
  \centering
    \includegraphics[width=19cm,trim={2.5cm 0 0.5cm 0cm},clip]{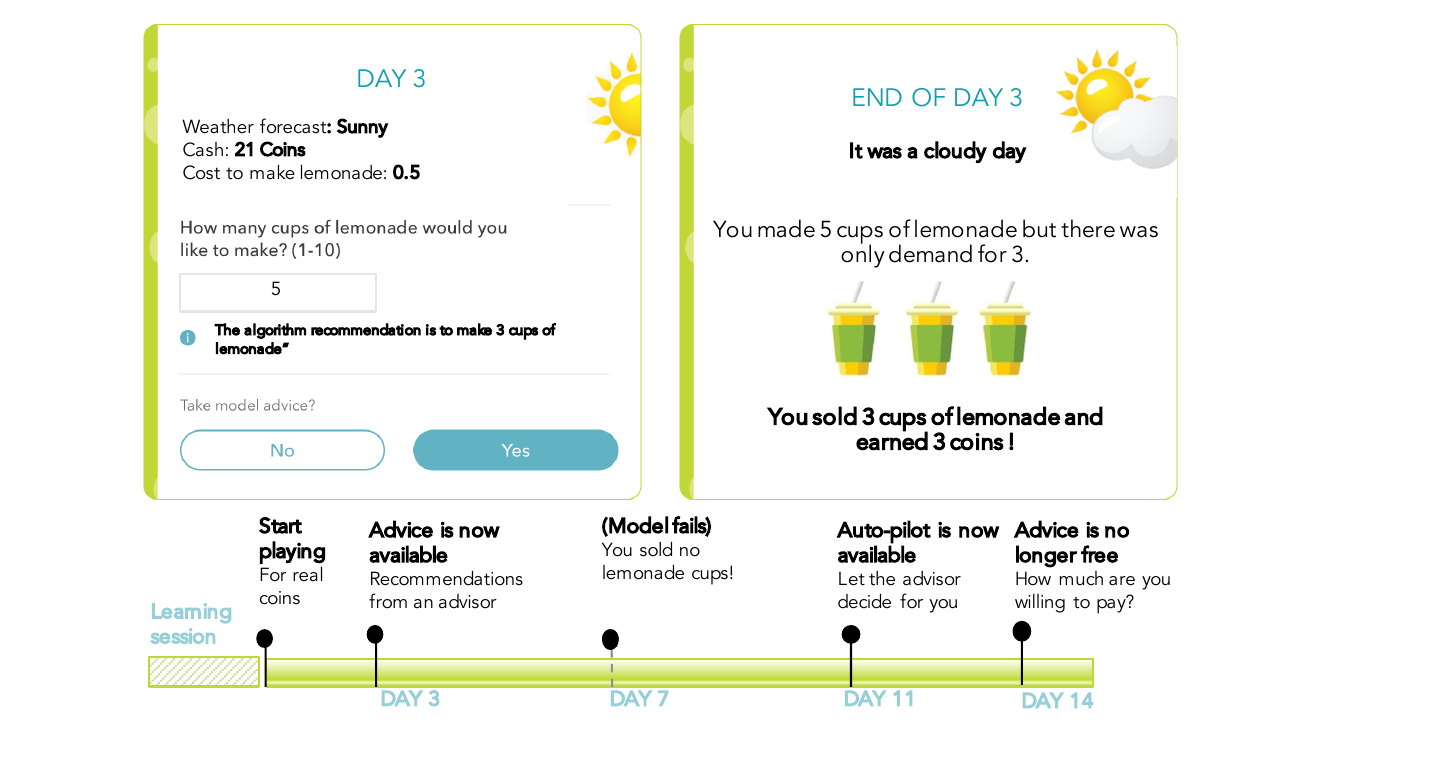}
    \caption{Illustration of one day at the lemonade stand (top), and the general flow of the lemonade stand game (bottom). Each day the player is given the weather forecast and asked how many cups of lemonade they would like to produce. After selecting the desired amount, advice is presented and the user can either accept the advice or produce the amount previously stated. Finally, the day is concluded with a screen that summarizes the events of the day, including the amount produced, demand, and earnings. }
  \label{fig:game_flow}
\end{figure*}
\subsection{Experimental Conditions}
\label{subsec:exp-conditions}

The game starts with a short learning session (3 days) designed to verify participants understanding. These days are not counted in the day numbering used throughout the analysis. After training, participants play the game for 14 game-days for real game coins. The complete game flow is described in the bottom graphic in Figure \ref{fig:game_flow}.

\textit{Initial adoption} -- At day 3, after completing 2 days of playing for real game money, participants were given the option to take recommendations from a so-called "advisor". Each recommendation by the advisor was accompanied with a short explanation which varied between experimental conditions (details below). To avoid an anchoring effect, recommendations were always displayed after participants already decided how many cups to make. They then could change their decision based on the recommendation (a \textit{"Take Advice"} button allowed them to switch to the number prescribed by the advisor. No other change to the the amount of lemonade production was allowed at this stage); alternatively, they could stay with their original choice. Initial adoption following the first impression was measured by the Readiness To Adopt (RTA) in this phase, the percent of users that took the advice each day, and was compared between the different explanation treatments (see table 2\ref{subsec:exp-conditions}).
\textit{Adoption gain} -- during the first days of the experiment (between days 3 and 7) the advice in all explanation treatments was set to perform perfectly (i.e. the advice given is precisely the actual demand that will occur). During this period we test how confidence in the algorithm advice is gained upon exposure to the accurate advisor, and specifically we compare the different conditions on day 7 after 4 rounds of exposure.

\textit{Advice failure and algorithm aversion} -- On day 7, The weather forecast was sunny for all participants, and the recommendation drastically fails, with a recommendation to produce the maximal amount of lemonade (10 cups) and actual demand for the minimal amount for a rainy day(0 cups). We explore the effect of experimental treatment (different explanations) on RTA the next day (i.e. day 8). From day 9 and until day 10 we look at the accumulation of renewed trust, as the algorithm returns to its accurate predictions. 

\textit{Automatic adoption} -- starting from day 11, \textit{"Auto-pilot"} mode is made available. Namely, participants are able to request the model to decide how many cups to make, without ever seeing its recommendations, or entering their decision manually (\textit{"Decide for me"}). By using this, users get the option not to think about amounts of production, and this makes sense for them if they know they are going to take the advice of the model anyway. The RTA here is measured as the number of participants who adopt the algorithm either automatically or manually as in the previous days.

\textit{Willingness to pay} -- During the last day of the game (day 13), participants are told that advice will no longer be free, and are asked how much they are will to pay to keep receiving recommendations in the following days. The purpose of this part is to check whether the adoption rate is affected by the algorithm not being free and to explore whether there are differences in the willingness to pay between the different mechanisms \cite{ben2020robo}.

Participants were randomly assigned to 5 experimental conditions, which differ only in the explanations for the advice that is given (See Table \ref{tab:exp-conditions} for the full text of the explanations provided for each group). Each subject participated was consistently shown only one type of advice throughout the experiment. 
Participants in the \textit{Human Expert} group were informed that advice is being generated by a so called human expert in the field of lemonade stand operations. This condition was designed to asses whether the mere usage of a computer algorithm has an effect on trust and adoption, and whether there is an algorithmic aversion as previously found by \cite{dietvorst2015algorithm}. In all other groups, participants were told that the suggestions are based on a so called computer algorithm. In the \textit{No Explanation} condition, participants were given no information about how the algorithm generates the predictions. This condition was used as baseline against which all other algorithm explanation treatments are compared. 

In the \textit{Global Explanation} condition, participants were given very general information about how the model operates, namely the type and extent of data it uses (sales data from many lemonade stands over several years). This condition simulates a very prevalent real-world scenario where we aim to increase model transparency with minimal cost in computation time and resources. In the \textit{"Feature-based"} condition, on the other hand, participants were given a fully detailed account about the features that were used to generate the specific prediction, similar to using local explanations in a real world scenario. To test the additive value of this information, it was attached to the baseline global explanation. Finally, to test the potential effect of reporting model performance, in the \textit{Performance-Based} condition information regarding performance ("with 90 percent certainty") was also attached to the baseline global explanation (we note that the performance observed by participants during the experiment is indeed statistically consistent with the 90 percent accuracy value mentioned to them). 

To clarify, all participants faced the same game phases and process with only one difference, the label on the advice offered. Given that the only difference was the "label" of the advice, and the fact that participants were randomly assigned to the conditions, no significant difference in RTA and WTP should be observed.
Any observe differences serve as an indication of prior perceptions and preferences relates to the given label.


\subsection{Participants}
The study was conducted via the online Amazon Mechanical Turk platform (AMT). The total time participants had to complete the experiment was 25 minutes. The average time of experiment completion were 12.66 Minutes. There were initial 514 respondents to the experiment. We decided to add a screening mechanism and added a minimal length of time of 6 minutes for completing the experiment in order to filter out individuals who did not play the game with sufficient attention and engagement.\footnote{Based on our simulations, it seems that a lower experiment duration is not feasible by an attentive individual.} 65 participants who finished in non-realistic time of less than 6 minutes were excluded. Our final sample consist of 449 subjects. These were randomly distributed between five explanations conditions, 84 of the participants received the \textit{Human Expert} treatment, 104 received the Algorithm with \textit{No Explanation}, 89 of the participant were allocated to the algorithmic advice with \textit{Global Explanation}, 87 received the \textit{Feature-based} explanation the the rest of 85 participants received the \textit{Performance-Based}. All participants were located in the U.S., with an average age of 36.6 (Std 11.2), 58.8\% Male, 68\% report working full time, 21.3\% part time, and the rest unemployed. 55\% of our participants report having an academic degree, 12\% an associate degree, and the rest a high school education or below.

\section{Results}
\begin{figure*}[t]
  \centering
    \includegraphics[width=\linewidth]{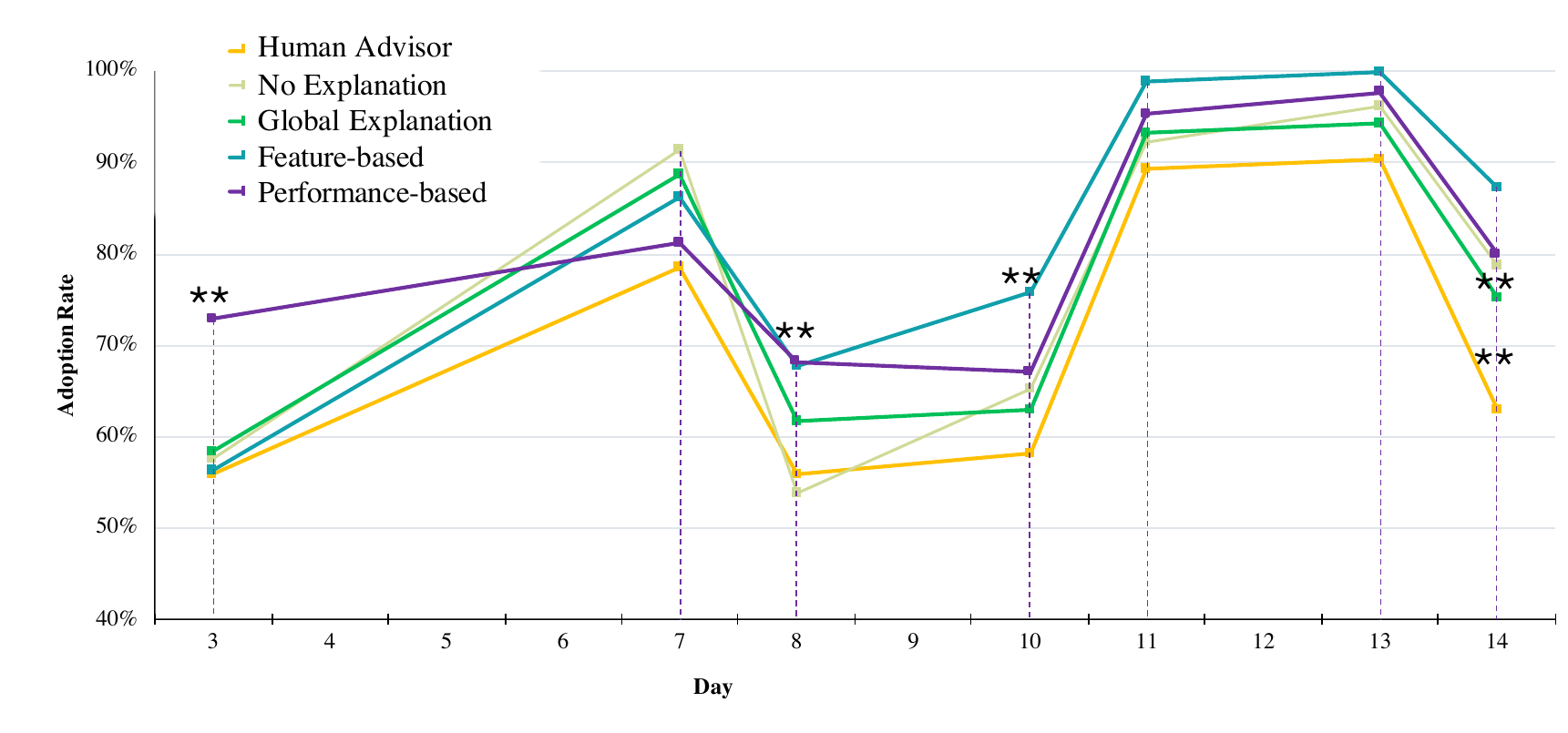}
    \caption{The average acceptance rate of model advise per game-day in each of the treatment groups. See Figure \ref{fig:game_flow} for the description of the flow of the game. }
  \label{fig:adoption}
\end{figure*}
\subsection{Evolution of Adoption}

We compare the Readiness to Adopt (RTA) for the different experimental condition groups throughout the experiment "two weeks" game duration. We start by describing the results with descriptive statistics and t-tests, followed by providing a Probit multivariate model regressions on the RTA while controlling for several relevant variables such as gender, age, trust and explanation perceived goodness indexes as measured by the questionnaires. Regression tables include the exact values for statistics reported throughout this section. 

Figure \ref{fig:adoption} summarizes the overall adoption rate (RTA) for different experimental conditions. The overall model adoption changed throughout the game, supporting our reference-dependent approach. On the very first day when the model was introduced (day 3) we see an average of 60\% adoption when averaging over all participants and all experimental conditions. Over the next few days, during which the algorithmic performance was good, overall adoption increased leading to an average adaption of 85.5\% on day 7. Following the model failure on day 7, adoption plummeted, returning approximately to the initial level with 61\% on day 8, this was followed by a slow recovery up until day 10 (66\%). With the introduction of the \textit{auto-pilot} option on day 11 adoption soared and reached the peak value of 94\%. Finally, when participants were told advice will no longer be free, and were asked about willingness to pay, we see an overall decrease in adoption on the final day (77\%).

Next, we break down the effect of the experimental explanation treatments and provide results of the RTA for each group. 

\begin{table*}[t]
    \centering
\begin{tabular}{lllllll}
\hline
                            & Day 3   & Day 7    & Day 8   & Day 10  & Day 13   & Day 14  \\
\hline
Performance-Based        & 0.343*  & -0.452*  & 0.373*  & 0.108   & 0.19     & -0.0317 \\
                            & (0.195)  & (0.239)   & (0.192)  & (0.194)  & (0.392)   & (0.213)  \\
                            &         &          &         &         &          &         \\
Feature-Based         & -0.106  & -0.286   & 0.354*  & 0.367*  & 0        & 0.229   \\
                            & (0.189)  & (0.247)   & (0.193)  & (0.2)    & (.0)      & (0.227)  \\
                            &         &          &         &         &          &         \\
Global Explanation          & -0.0324 & -0.147   & 0.222   & -0.0109 & -0.28    & -0.168  \\
                            & (0.189)  & (0.253)   & (0.188)  & (0.192)  & (0.335)   & (0.208)  \\
                            &         &          &         &         &          &         \\
Trust index                 & 0.0976  & 0.103    & 0.376** & 0.331** & 0.15     & 0.153   \\
                            & (0.15)   & (0.191)   & (0.151)  & (0.156)  & (0.318)   & (0.171)  \\
                            &         &          &         &         &          &         \\
Explanation\_goodness index & 0.374** & -0.159   & -0.0393 & -0.304* & 0.288    & 0.274   \\
                            & (0.153)  & (0.193)   & (0.154)  & (0.159)  & (0.336)   & (0.175)  \\
                            &         &          &         &         &          &         \\
constant                    & -0.154  & 1.543*** & 0.173   & 0.612*  & 2.215*** & 0.680** \\
                            & (0.31)   & (0.39)    & (0.307)  & (0.328)  & (0.608)   & (0.342)  \\
                            &         &          &         &         &          &         \\
N                           & 365     & 365      & 365     & 365     & 278      & 365 \\
\hline
\end{tabular}

\caption{Notes: Probit regressions, robust standard errors are in parentheses. The dependent variable is the RTA over the different days. The Algorithmic with No explanation is the underlying condition where other treatments are dummy variables. Trust index is a dummy variable that receives 1 for individuals whose index score of all seven trust questions (see appendix 2 for the trust questions) is above the median; Explanation goodness index is a dummy variable that receives 1 for individuals that their index score of all perceived explanation goodness index questions is above the median (see appendix 2 for the perceived explanation goodness questions). The estimation includes constant and control vector X that consist of age, gender, and age exponent  ("$* p < 0.10 ** p<0.05 *** p<0.01$").}
\end{table*}

\begin{table*}[t]
\begin{tabular}{lllllll}
\hline
    & Gained Trust & Lost Trust & Recovery & Auto Advice & No Longer Free &  Tobit - WTP           \\
\hline
Performance-Based        & -1.240***    & -0.501**   & 0.0986   & 0.492                 & 0.149          & 0.840**     \\
                            & (0.458)       & (0.214)     & (0.317)   & (0.567)                & (0.217)         & (0.344)      \\
                            &              &            &          &                       &                &             \\
Feature-Based        & -0.519       & -0.419**   & 0.426    & 0                     & -0.115         & 0.537*      \\
                            & (0.454)       & (0.208)     & (0.312)   & (.0)              & (0.23)          & (0.318)      \\
                            &              &            &          &                       &                &             \\
Global Explanation          & -0.606       & -0.262     & 0.0886   & -0.0565               & 0.166          & 0.788**     \\
                            & (0.459)       & (0.201)     & (0.299)   & (0.453)                & (0.215)         & (0.38)       \\
                            &              &            &          &                       &                &             \\
Human\_advisor              &              &            &          &                       &                & 0.454       \\
                            &              &            &          &                       &                & -0.286      \\
                            &              &            &          &                       &                &             \\
Trust index                 & -0.0615      & -0.323**   & 0.628**  & -0.0154               & -0.101         & -0.046      \\
                            & (0.317)       & (0.162)     & (0.266)   & (0.481)                & (0.175)         & (0.241)      \\
                            &              &            &          &                       &                &             \\
Explanation\_goodness index & -0.613*      & 0.15       & -0.417   & 0.57                  & -0.337*        & 0.213       \\
                            & (0.325)       & (0.168)     & (0.269)   & (0.486)                & (0.179)         & (0.258)      \\
                            &              &            &          &                       &                &             \\
N                           & 135          & 318        & 137      & 97                    & 365            & 449        \\
\hline
\end{tabular}
\caption{Notes: columns 1-5: Probit regressions, robust standard errors are in parentheses. The dependent variable is the RTA differences over the experiment phases. Gained Trust - RTA day 7 adopters minus RTA day 3, Lost Trust - RTA day 7 adopters minus RTA day 8 ,Recovery -  RTA day 10 adopters minus RTA day 8 , Auto Advice Available - RTA day 13 minus RTA day 11, No Longer Free - RTA day 14 minus RTA day 13. The Algorithmic with No explanation is the underlying variable where other treatments are dummy variables. Trust index is a dummy variable that receives 1 for individuals whose index score of all seven trust questions (see appendix 2 for the trust questions) was above the median;  Explanation Goodness index is a dummy variable that receives 1 for individuals that their Index score of all perceived explanation goodness index questions is above the median (see appendix 2 for the perceived explanation goodness questions). Column 6: Tobit regression, robust standard errors are in parentheses. The dependent variable is the WTP on the last day of the game. Other variables are as columns 1-5. The estimation includes constant and control vector X that consist of age, gender, and age exponent ("$* p<0.10 ** p<0.05 *** p<0.01$").}
\end{table*}

\subsection{Algorithm aversion (Human expert v.s. Algorithmic advice)}
First, we examine the effect of human versus algorithmic origin of advice with no further explanation. Results show no evidence of algorithm aversion-- namely there is no significant difference in the initial adoption of the \textit{Human Advisor} (RTA of 56\%) as opposed to the computer algorithm \textit{No Explanation} conditions (with an RTA of 57.7\%). However, over the first few days, the algorithmic advice with \textit{No Explanation} showed significantly gains in adoption as compared to the \textit{Human Advisor} leading to a 16.3\% percent gap on day 7 (P-value = 0.0129, t = 2.5114). Following model failure on day 7, adoption and RTA levels fell back to around the initial values in both cases (54\% and 56\% for \textit{No explanation} and \textit{Human Advisor}, respectively). Recovery in the next few days is slightly and not significantly more favorable for the AI-based advisor (65\% compared to 58\% by day 10), and this trend remains throughout the experiment. The decline in adoption following the probe of willingness to pay is significantly larger in the Human Advisor condition with 79\% compared to 63\% for algorithmic \textit{No Explanation} and the \textit{Human advisor}, respectively (P-value = 0.0169, t = 2.4104). 

For the regression analysis we proceed by conducting a Probit multivariate model regressions. The base specification is

\begin{equation}
    RTA_{i, t} = \alpha + \beta  T_{i, t} + \delta \cdot \textbf{T\_E}_{i, t} + \gamma \cdot \textbf{X}_{i, t} + \epsilon_i  
\label{eq:model-1}
\end{equation}


where RTA is the outcome variable associated with individual $i$ with time difference of $t$ game "days" of the experiment. T is a dummy variable for the type of investment advice: "Human Advisor" (where the underlying condition is the "AI-No explanation" alternative). $T\_E$  is a vector of the Trust and Explanation satisfaction dummies for indicating if an individual is above the median on the index based on questionnaires developed by \cite{gefen2003trust,komiak2006effects,mcknight2002developing,hoffman2018metrics}. 
\footnote{The questionnaires can be found in appendix A} X is the vector of age and gender variables. Table 1 shows the Probit analysis results.
We observe similar results when including the range of the control variables. For detailed regression specification, see Appendix A. 


The absence of algorithm aversion observed in our study may be the result of using the Mturk platform, an online platform that may lead to the more technological orientation of individuals using it, compared to the general population.

\subsection{The effect of explanation type on RTA}
Next, we considered the effect of the type of explanations provided by the algorithmic advisor on RTA during the different phases of the game. As before, we start with descriptive statistics and continue with a multivariate regression analysis. For the analysis, we are conducting a Probit,  \footnote{The results are not sensitive to the choice of regression models such as OLS or Logit.} cross-section multivariate model regressions, on the different treatments while controlling for gender, age, trust antecedents index and Explanation satisfaction index with two approaches: (A) between conditions analysis as in model (\ref{eq:model-1}) on the different experiment days, and (B) on the change in adoption between the different experiment phases. The base specifications are:
\begin{equation}
    RTA_{i, t} = \alpha + \beta  T_{i, t} + \delta \cdot \textbf{T\_E}_{i, t} + \gamma \cdot \textbf{X}_{i, t} + \epsilon_i  
\label{eq:model-2}
\end{equation}
\begin{equation}
    RTA_{i, t}^{\Delta} = \alpha + \beta  T_{i, t} + \delta \cdot \textbf{T\_E}_{i, t} + \gamma \cdot \textbf{X}_{i, t} + \epsilon_i  
\label{eq:model-3}
\end{equation}

For model (\ref{eq:model-2}), the outcome variable is the RTA for individual i in the different $t$ "days" of the experiment and the rest is identical to the specified above in model (\ref{eq:model-1}). Table 3 shows the Probit analysis results. For model (\ref{eq:model-3}), the analysis is done on the change in adoption between the experiment phases. The outcome variables are as follows: RTA day 7 minus RTA day 3 (as gaining trust), RTA day 7 minus RTA day 8 (as loss of trust), RTA day 10 minus RTA day 8 (as a Recovery), RTA day 13 minus RTA day 11 (as  auto advice available), RTA day 14 minus RTA day 13 (as willingness to pay). The explanatory variables are similar to as specified above in model (\ref{eq:model-1}). Table 4 shows the Probit analysis results for these models.

\subsubsection{Initial adoption (day 3)}
When comparing the \textit{Global Explanation} with \textit{No explanation} we find no significant difference in initial adoption. When we attach to the Global explanation the more detailed feature-Based explanation, we again find no significant difference. However, attaching the accuracy of the model to the explanation created a significant difference and created the highest initial adoption rates, with RTA of 73\% for the Accuracy compared to an RTA average of 57\% for the other explanation types (P-values = 0.0293, t = 2.1963). From the regression with controls, we find similarly that adding the explanation of \textit{Performance-Based} had significantly more adoption compare to \textit{No-explanation}. In addition, when \textit{Explanation satisfaction} was perceived as high (above median), the initial adoption was significantly higher as well.

\subsubsection{Gaining confidence and trust (day 7)}
During these first four days, the advice provided in all experimental conditions allowed participants to make the highest possible profit (i.e. perfect predictions and advice). In line with this, all advice treatment alternatives gained user confidence, and adoption raised significantly by an average of 42.5\%. When comparing the different experimental conditions, we find no significant differences except the \textit{Performance-Based} explanation which gained less adoption compare to the no explanation alternative (P-value = 0.0403, t = 2.0652). From the regressions with controls, we find, similarly to above, that only the \textit{Performance-Based} gained less adoption compare to the no-explanation. This result was found in both regression approaches. These results suggest that when advice performance is immaculate, the type of explanation presented with it is less important to individuals.  

\subsubsection{Advice fails and algorithm aversion (days 8-10)}
Following algorithm failure on day 7, we find an average adoption drop of 28\% on day 8. When we explore the different explanations we find the following: the two more elaborate explanations, the \textit{Performance-Based} (P-value = 0.0445, t =  -2.0225) and the \textit{Feature-Based} (P-value =0.0498, t =  -1.9741) prove to be more resistant to the effect of algorithm error. The adoption drop was significantly lower compared to the \textit{No explanation} alternative with 16\% and 21\% reduction for the \textit{Feature-Based} and the \textit{Performance-Based} ,respectively and 41\% reduction for the advice with No explanation. We find no significant differences between the \textit{Global Explanation} and the other alternative explanations. When we explore the adoption recovery, during the next few days and until day 10, we find that the \textit{Performance-Based} explanation had a marginally significant (P-value = 0.0633, t = 1.8691) tendency to recover better than the global explanation (12\% compared to only 2\% for the \textit{Feature-Based} and the \textit{Global Explanation}, respectively). 

In the regressions with controls, we find both in model (\ref{eq:model-2}) and (\ref{eq:model-3}) that the \textit{Feature-Based} and \textit{Performance-Based} have significantly lower adoption drop compare to the alternative with No explanation. In addition, when the trust was perceived as higher, it lowered the drop rates and increased the recovery from the failure. Lastly, on day 10, we observe a marginally significant higher adoption rate for the \textit{Feature-Based} condition compare to the No-explanation alternative.

\subsubsection{Auto take advice and trust (days 11-13)}
On day 11, subjects have an option to select auto advice before starting the day. The new instrument that enables the participants to avoid deciding by themselves has a major impact. We observe adoption jump of 40\% on average to a mean RTA of 95\% on day 11 and up to 97\% on day 13 (For the algorithmic treatment alternatives that included explanations (Global, Feature-Based, Performance-Based). Among the alternative explanations, we find that the \textit{Feature-Based} explanation achieved the highest adoption rates and that they are marginally significant compare to the No explanation alternative (P-value = 0.0651, t = 1.8557) and significantly compared to the \textit{Global} explanation (P-value = 0.0249, t = 2.2627).
In the regressions, we find similarly to above, no significant differences between the experimental conditions to the no explanation alternative. 

Concerning the adoption jump, we relate our results to the fact that making financial decisions is a challenging task that people reluctantly enter \cite{gennaioli2015money}. The way the advice is now offered enables the participants to transfer the responsibility for the decision making to the advisor and avoid making it by themselves. Also, the relatively low stakes of the game (winning several U.S dollars at the end of the experiment) can serve as an explanation of why participants transfer responsibility and avoid effort.
\subsubsection{Advice is no longer free (day 14)}
On the last day we introduced a new situation where the advisor is no longer free and asked participants whether they want the advice and if so, how much they are willing to pay for it. During this round we observe adoption drop of 21\% for the algorithmic alternatives. This was despite the fact that we allowed the subjects to self-determine the value (the price) of the advice. The \textit{Feature-Based} condition showed a tendency for lower adoption reduction (only 14.5\% RTA drop). This result was significant compare to the \textit{Global Explanation} (P-value = 0.0398, t =  -2.0712). No other significant difference were observed between the other algorithmic alternatives.
From the regression analyses, we find, no significant differences between the experimental conditions to the no explanation alternative.
\subsection{Willingness to pay}
The Willingness to Pay measurement, as specified above, can further attest to way participants perceive the advice given by the different advice mechanisms. We find that participants assigned to the No-explanation alternative were willing to pay the least for the use of the advice with mean payment of 1.005 game coins, compared to a mean payment of 1.774 game coins for all other "AI advice" alternatives -- A 76.5\% gap. The difference in willingness to pay was significant for \textit{Global Explanation} and \textit{Performance-Based} conditions ((P-value = 0.0150, t = 2.4535), (P-value = 0.0063, t = 2.7621), respectively) and marginally significant for \textit{Feature-Based} condition (P-value = 0.0566, t = 1.9181). No significant differences were observed between the other alternatives. In addition, we observe positive marginal significant difference in the willingness to pay for the Human-expert compared to the No-explanation ((P-value = 0.0611, t = 1.8836). This result is in contrast to what we observe with respect to the RTA over the same day.

For the regression analysis, we conduct a Tobit, \footnote{The results are not sensitive to OLS regression} cross-section multivariate model regressions, on the different treatments while controlling for gender, age, trust antecedents index, and Explanation satisfaction index. The base specifications is:

\begin{equation}
    WTP_{i} = \alpha + \beta  T_{i} + \delta \cdot \textbf{T\_E}_{i} + \gamma \cdot \textbf{X}_{i} + \epsilon_i  
\label{eq:model-4}
\end{equation}

The outcome variable is the Willingness to Pay (WTP) for individual i. T is a dummy variable for the type of investment advice: \textit{Global Explanation} and \textit{Performance-Based} and \textit{Feature-Based} and \textit{Human-Expert} (where the underlying condition is the "AI-No explanation" alternative). The rest is identical to the specified above in model (\ref{eq:model-1}). Regression analysis results are as described above and specified as part of model (\ref{eq:model-4}) in Table 4 (columns 6).


From the regression analyses, we find that the willingness to pay for \textit{Global Explanation} and \textit{Performance-Based} conditions were significantly different and positive compared to the \textit{No-explanation} alternative as observed by the t-tests and the same marginal significant tendency toward higher payment for \textit{Feature-Based} condition compare to the \textit{ No-explanation}. In addition, we find a significant age U shape for the willingness to pay. Consisting with \cite{ben2020robo} results.  
For the comparison of the Human-Expert vs. No-Explanation, we find that with the controls, the effect observed is not significant. Meaning that we don't find a willingness to pay significant differences between those advice alternatives.

Our results imply that adding explanations increases the willingness to pay for the advice. Also, it seems that using "keep it simple and short" explanations with \textit{Global or Performance-Based}, had the highest impact on the participant's willingness to pay. We refer our results to \cite{gennaioli2015money} theory in which people are willing to pay more for an advisor that they trust more. It seems that adding WPT in addition to RTA had an additional dimension of trust relating to algorithmic advice. 

\subsection{Explanations, trust, and satisfaction}

To reinforce our measure of adoption, and validate the relation to trust, we examined the effect of trust antecedents questions which were found as relevant in the literature \cite{gefen2003trust,mcknight2002developing,komiak2006effects}, on the adoption of the advice alternatives using within treatment analysis, over the various days. We used a validated index including 7 trust questions as the measure. We find that on day 8 and day 10, after the algorithm failed, individuals who indicated that they trust the advice more indeed adopt it more. In particular, individuals who trusted the advice more with \textit{Feature-Based} explanation and \textit{Performance-Based} based explanation by our index, showed higher adoption rates. In addition to the trust index, when we explore the explainable goodness questionnaire index, we find that it was significantly influential on day 3 adoption. Individuals who valued the explanation more, adopt the advice more. Regression analysis results are as described above and specified as part of model (\ref{eq:model-2}) and (\ref{eq:model-3}).

\section{Conclusion and Discussion}

Machine learning has become a prevalent technology, especially over the last decade, with applications and implications in many aspects of everyday life. In order to promote a much needed cooperation between human and machine, one of the research goals in the field of explainable AI is to find the most successful ways of explaining and presenting AI-based decisions to individuals. The majority of this line of work is focused on algorithmic methods to produce explanations for complex machine learning methods. However, the complimentary question of whether different types of explanations are important to create trust and adoption is for the most part still open. To this end, we constructed a decision making game framework to test adoption, trust, and willingness to pay for financial machine learning models. Unlike questionnaire based paradigms, the \textit{lemonade stand game} allows us to test these issues quantitatively, based on actual behavior in the presence of real financial stakes.

To the best of our knowledge, this is the first study to directly evaluate the behavior of users in response to varying types of textual global and local explanations of AI over time in different situations. Our experimental paradigm allows testing initial adoption in several experimental conditions (defined by the different types
of explanations), as well as the evolution of the relations over time when the AI advice proves to be useful, or after it fails. Post-game questionnaires further allow integration of measures common in the behavioral sciences for additional validation of our findings.

We find that attaching different explanations created a significant difference. We observed no algorithm aversion, and, in our experiment, participants were more inclined to adopt so-called AI-given advice compared to so-called Human advice. We find that an accuracy-based explanation of the model in initial phases led to higher adoption rates. When the performance is immaculate, there is less importance associated with the kind of explanation for adoption. In addition, in cases of failure, using more elaborating feature-based or accuracy-based explanations helps substantially in reducing the adoption drop, and the negative influence on trust towards the algorithm. 

Furthermore, using an autopilot increased adoption significantly, a finding which we relate to the unwillingness to make decisions if possible. Presenting a feature-based explanation partially mitigate the adoption drop caused by the question about willingness to pay. Participants which were assigned to "AI-advice" with explanations were willing to pay more for the advice compare to the No-explanation alternative. We find a correlation between trust antecedents and our measurement of RTA. When levels of trust were high, we observed mitigation in adoption drop when failure occurs, and a correlation with gaining recovery after this failure. We observe positive correlation between Explanation satisfaction perceptions with initial adoption levels as well. Lastly, we find that age is negatively correlated with auto-take advice and a tendency (not significant) of age and lower adoption rates. 

We contribute to the literature in three key ways: 

(1) First, we document that there is no single best explanation that fits all. The type of explanation that is best suited to promote trust is time and situation dependent. Namely, the "What" we should explain depends on the "when". Interestingly, participants were more inclined overall to adopt AI-given advice than follow a human expert. The effect of the model (and human advice) failure on subsequent advice taking is remarkable, with a single failure causing adoption rates to plunge. 

(2) Second, our results suggest that often the end-user doesn't need to know more than very general facts to accept the system. Our study shows that such general explanations yield good results in terms of RTA and WTP compared to the alternative of no explanations, and especially after AI failure. Moreover, stating accuracy statistics about the algorithm can further strengthen the aforementioned effect. Furthermore, explanations increased the participants willingness to pay for the advice. Results of the current study highlight the potential utility of simple explainability solutions in these aspects. 

(3) Third, our experimental paradigm of the lemonade stand game applies the RTA measurement in different and evolving situations into the explainable AI literature. This framework may be utilized to answer other questions in the field of explainable AI and trust in AI. 
Future work can investigate the effect of different explanation types in broader conditions such as, but not limited to: varying reported accuracy values of the algorithm and over more extended periods. Whether Willingness to Pay changes over time with different situations? Will using additional experimental methods such as a more controlled lab experiment or a field experiment on a live system show the same results? 
Some of the key components can be implemented in research studies on different explainability techniques, such as visual explanations. Answering these questions in addition to our study will help bridge the gap between the immense technical success we see in machine learning, and the difficulty often faced when trying to understand and amplify adoption and trust among users.

\bibliographystyle{ACM-Reference-Format}
\bibliography{bibliography.bib}
\clearpage

\includepdf[pages=-]{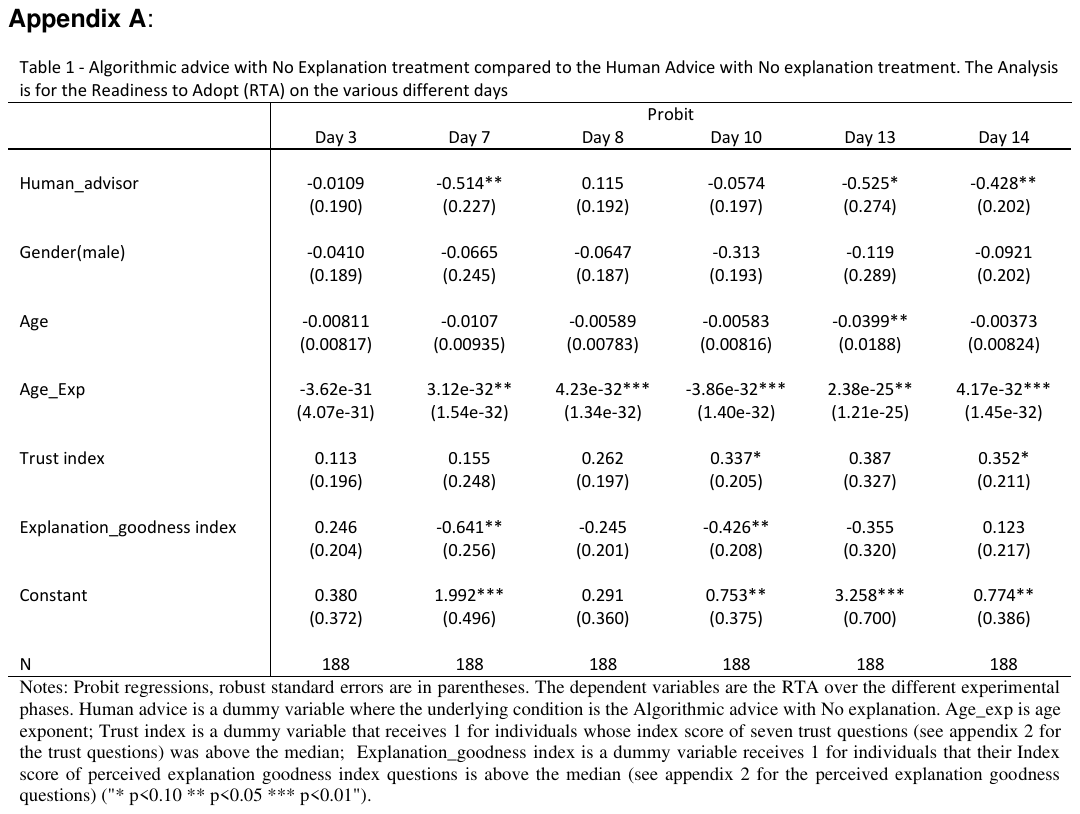}

\end{document}